\documentclass[superscriptaddress,twocolumn,showpacs,preprintnumbers,amsmath,amssymb,prl]{revtex4}

\usepackage{graphicx}
\usepackage{dcolumn}
\usepackage{bm}
\usepackage{multirow}
\usepackage{color}

\begin{document}

%\preprint{APS/123-QED}

\title{Incommensurate Magnetism in FeAs Strips: Neutron Scattering from CaFe$_4$As$_3$}

\author{Yusuke Nambu}
\affiliation{Institute for Quantum Matter and Department of Physics and Astronomy, The Johns Hopkins University, Baltimore, Maryland 21218, USA}
\affiliation{NIST Center for Neutron Research, National Institute of Standards and Technology, Gaithersburg, Maryland 20899, USA}
\author{Liang L. Zhao}
\affiliation{Department of Physics and Astronomy, Rice University, Houston, Texas 77005, USA}
\author{Emilia Morosan}
\affiliation{Department of Physics and Astronomy, Rice University, Houston, Texas 77005, USA}
\author{Kyoo Kim}
\affiliation{Department of Physics, Rutgers University, Piscataway, New Jersey 08854, USA}
\author{Gabriel Kotliar}
\affiliation{Department of Physics, Rutgers University, Piscataway, New Jersey 08854, USA}
\author{Pawel Zajdel}
\affiliation{Institute of Physics, University of Silesia, Uniwersytecka 4, 40-007 Katowice, Poland}
\author{Mark A. Green}
\affiliation{NIST Center for Neutron Research, National Institute of Standards and Technology, Gaithersburg, Maryland 20899, USA}
\affiliation{Department of Materials Science and Engineering, University of Maryland, College Park, Maryland 20742, USA}
\author{William Ratcliff}
\affiliation{NIST Center for Neutron Research, National Institute of Standards and Technology, Gaithersburg, Maryland 20899, USA}
\author{Jose A. Rodriguez-Rivera}
\affiliation{NIST Center for Neutron Research, National Institute of Standards and Technology, Gaithersburg, Maryland 20899, USA}
\affiliation{Department of Materials Science and Engineering, University of Maryland, College Park, Maryland 20742, USA}
\author{Collin Broholm}
\affiliation{Institute for Quantum Matter and Department of Physics and Astronomy, The Johns Hopkins University, Baltimore, Maryland 21218, USA}
\affiliation{NIST Center for Neutron Research, National Institute of Standards and Technology, Gaithersburg, Maryland 20899, USA}

\date{\today}

\begin{abstract}
Magnetism in the orthorhombic metal CaFe$_4$As$_3$ was examined through neutron diffraction for powder and single crystalline samples.
Incommensurate (${\bm q}_{\rm m}\approx (0.37-0.39)\times{\bm b}^{\ast}$) and predominantly longitudinally ($\parallel b$) modulated order develops through a 2nd order phase transition at $T_{\rm N}=89.63(6)$ K with a 3D Heisenberg-like critical exponent $\beta=0.365(6)$.
A 1st order transition at $T_2=25.6(9)$ K is associated with the development of a transverse component, locking ${\bm q}_{\rm m}$ to $0.375(2){\bm b}^{\ast}$, and increasing the moments from 2.1(1) to 2.2(3) $\mu_{\rm B}$ for Fe$^{2+}$ and from 1.3(3) to 2.4(4) $\mu_{\rm B}$ for Fe$^+$.
The {\it ab-initio} Fermi surface is consistent with a nesting instability in cross-linked FeAs strips. 
\end{abstract}


\maketitle

The recent discovery of superconductivity (SC) in iron pnictides \cite{rev} has opened a new chapter in SC research.
In contrast to the cuprates, which display SC near a correlated insulating state, iron SC appears when doping or applying pressure to metallic parent compounds with a spin density wave (SDW).
A square, puckered iron chalgogenide or iron pnictide plane is the common theme in this family of materials, which so far has been realized in the ZrCuSiAs (1111), ThCr$_2$Si$_2$ (122), anti-PbO (11) and Cu$_2$Sb (111) structures.
Chemically related CaFe$_4$As$_3$ \cite{Liang,JACS} contains interpenetrating FeAs strips, which --as spin-ladders in copper oxides-- may provide unique insight into electronic correlations of the square lattice plane. 

Similar to the parent compounds of the 1111- and 122-type superconductors, CaFe$_4$As$_3$ is not superconducting down to 1.8 K \cite{Liang}.
The novel framework structure is composed of shared FeAs tetrahedra with Ca atoms in the channels so defined (Fig. 1(a)).
The structure is reminiscent of {\it porous} manganese oxides \cite{MnO2}.
Extending along the $b$-axis are FeAs strips where 4-fold As coordinated Fe (Fe1-Fe3 in Table I) occupy an approximately square lattice of finite width.
Fe atoms that link strips (Fe4 in Table I) on the other hand form FeAs$_5$ pyramids.
M\"ossbauer spectra \cite{JACS} indicate different electronic states on 4- and 5-fold coordinated sites consistent with formal valences of Fe$^{2+}$ and Fe$^{1+}$, respectively.
\begin{figure}[b!]
\includegraphics[width=210pt]{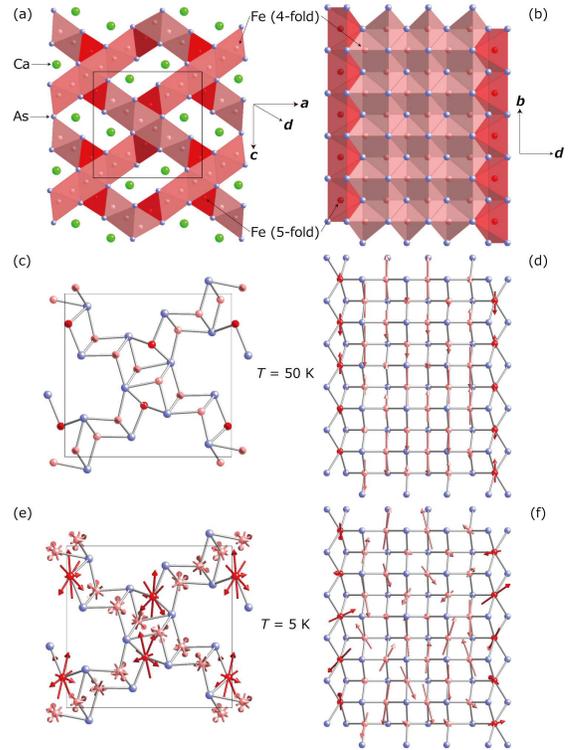}
\caption{(color online). Crystal (a,b) and magnetic structures of CaFe$_4$As$_3$ at (c,d) $T=50$ K and (e,f) 5 K in $ac$ plane (left) and along $b$-axis (right). Solid squares indicate a unit cell. The vector $\hat{\bm d}$ ($\hat{\bm a}\cdot\hat{\bm d}=\cos 30^\circ$) is used for right panels.}
\end{figure}

Two phase transitions were detected through thermal anomalies in bulk properties \cite{Liang,JACS}.
While only the upper transition at $T_{\rm N}=89.63(6)$ K is marked by a significant specific heat anomaly, the lower transition at $T_2=25.6(9)$ K yields a strong reduction in resistivity ($\rho$).
The magnetization is anisotropic with larger values along strips ($\parallel b$).
The large Sommerfeld constant ($\gamma=0.02$ J/mol-Fe/K$^2$) indicates relatively strong electronic correlations. 

To elucidate the magnetic properties of CaFe$_4$As$_3$, we carried out neutron scattering measurements on powder and single crystalline samples.
We show that the upper transition is to a longitudinal incommensurate (IC) SDW, while the lower transition is associated with the development of a transverse component and lock-in to commensurate order. 
A powder sample and a single crystal of CaFe$_4$As$_3$ were prepared according to Ref. \cite{Liang}.
Neutron powder diffraction data were collected on BT1 and BT9 at NIST.
Single crystal experiments were performed on BT9 and on the newly developed Multi Axis Crystal Spectrometer.
The error bars indicate $\pm$ the standard deviation.
We used group theoretical analysis to identify magnetic structures that are allowed by symmetry and consistent with the diffraction data.

We performed neutron powder diffraction measurements at $T=$ 5 K, 50 K, 100 K and room temperature (RT).
As in previous reports \cite{Liang,JACS}, the crystal structure is accounted for by space group $Pnma$. 
Table \ref{atomic} summarizes the RT structural parameters.
\begin{table}[t!]
\caption{Atomic positions within $Pnma$ of CaFe$_4$As$_3$ at RT determined by Rietveld analysis ($\chi^2=2.2$). Lattice constants are $a=11.9062(2)$ \AA, $b=3.74477(6)$ \AA, and $c=11.6131(2)$ \AA. Isotropic Debye-Waller factors ($U_{\rm iso}$) are employed.}
\label{atomic}
\begin{ruledtabular}
\begin{tabular}{lccccc}
Atom	&	Site	& $x$		& $y$		& $z$				& $U_{\rm iso}$ ({\AA}$^2$)	\\
\hline
As1	& $4c$	& 0.1364(4)	& $1/4$	& 0.0778(4)	& 0.0149(12)	\\
As2	& $4c$	& 0.3839(4)	& $1/4$	& 0.2636(4)	& 0.0093(13)	\\
As3	& $4c$	& 0.0998(4)	& $3/4$	& 0.3797(3)	& 0.0050(13)	\\
Fe1	& $4c$	& 0.1808(2)	& $1/4$	& 0.2763(2)	& 0.0055(9)	\\
Fe2	& $4c$	& 0.1941(3)	& $1/4$	& 0.8759(3)	& 0.0095(8)	\\
Fe3	& $4c$	& 0.4343(3)	& $1/4$	& 0.4635(3)	& 0.0064(9)	\\
Fe4	& $4c$	& 0.4770(3)	& $1/4$	& 0.6864(3)	& 0.0103(10)	\\
Ca	& $4c$	& 0.3376(4)	& $3/4$	& 0.0686(6)	& 0.0115(16)	\\
\end{tabular}
\end{ruledtabular}
\end{table}
Additional peaks associated with magnetic ordering were observed at 50 K and 5 K.
Figure 2 shows a portion of the Rietveld refined diffraction pattern, which is consistent with the single crystal data.
\begin{figure}[b!]
\includegraphics[width=175pt]{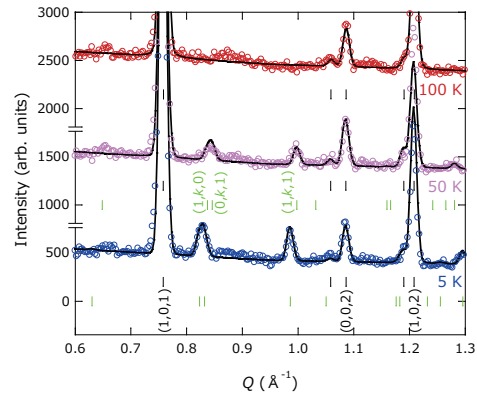}
\caption{(color online). A portion of the CaFe$_4$As$_3$ neutron diffraction pattern at 100 K, 50 K and 5 K with Rietveld refinement (solid lines). The black and green bars with representative ($hkl$) indices show the calculated positions of nuclear and magnetic peaks, respectively.}
\end{figure}
All the magnetic peak positions can be accounted for by a propagation wave vector, ${\bm q}_{\rm m}=(0k0)$.
The strongest magnetic peak actually consists of overlapping ($1k0$) and ($0k1$) peaks, where $k=$ 0.386(2) and 0.375(2) at 50 K and 5 K, respectively, indicating ${\bm q}_{\rm m}$ is temperature dependent.

To determine the temperature dependence of ${\bm q}_{\rm m}$, we used single crystal diffraction in the ($0kl$) plane on a $0.5\times 8\times 0.5$ mm$^3$ sample.
Figure 3 shows the $T$-dependence of the ($0k1$) magnetic peak.
The peak appears for $T\approx 90$ K along with an anomaly in $\rho(T)$ \cite{Liang}.
The temperature dependent integrated intensity ($I$) can be described as $I \propto t^{2\beta}$, where $t\equiv (T_{\rm N}-T)/T_{\rm N}$ is the reduced temperature and $T_{\rm N}=89.63(6)$ K (right inset to Fig. 3(b)).
The inferred critical exponent $\beta=0.365(6)$ is consistent with the 3D Heisenberg model, 0.36 \cite{AM}.

Below $T_{\rm N}$, the peak-center $k$ appears to increase continuously from 0.37 to 0.39 on cooling (Fig. 3(a)), which indicates an incommensurately modulated structure.
$I$ and $k$ then change abruptly near 25 K.
We associate this anomaly with a first order phase transition for the following reasons:
(i) There is considerable thermal hysteresis ($\Delta T\sim 2$ K) of $I$ and $\rho$ (Fig. 3(c)) and (ii) diffraction peaks associated with the two phases coexist close to $T_2$ (left insets to Fig. 3(b)).
Since no structural phase transition was detected at $T_2$, this appears to be an intrinsic first order magnetic phase transition.
This is different from observations in 122 systems, where a first order structural phase transition precedes \cite{122pre} or coincides \cite{rev} with a magnetic transition.
While the structural phase transition in 122 systems breaks tetragonal symmetry and relieves frustrating magnetic interactions such a scenario is not expected in orthorhombic CaFe$_4$As$_3$.
\begin{figure}[t!]
\includegraphics[width=180pt]{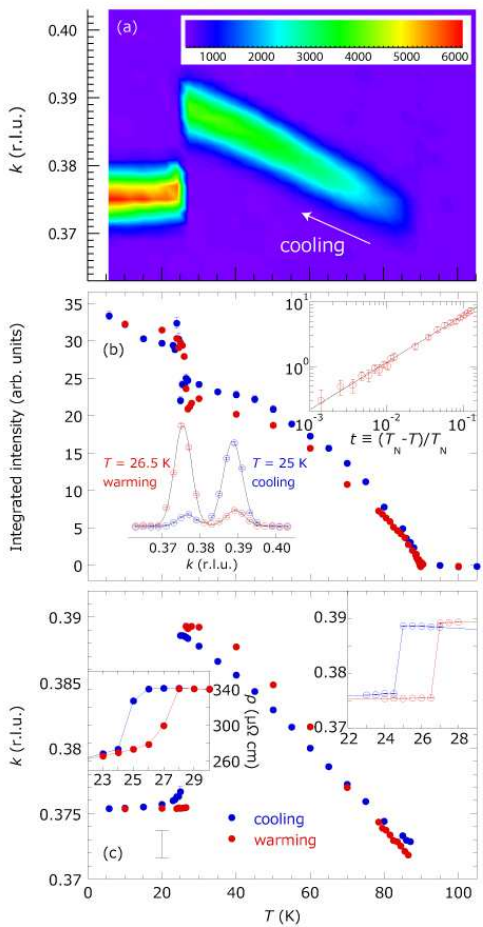}
\caption{(color online). (a) Color plot of the neutron scattering intensity versus temperature and wave vector transfer along the $(0k1)$ direction. The temperature dependence of the inferred integrated intensity and peak position are shown in frames (b) and (c) respectively. The vertical bar in (c) represents a systematic instrumental uncertainty. The insets to (b) show warming intensity against the reduced temperature near $T_{\rm N}$ and coexistence of two peaks near $T_2$. The insets to (c) emphasize thermal hysteresis of $I$ and $\rho$.}
\end{figure}

Below $T_2$, $k$ is $T$-independent and fixed at 3/8 within experimental accuracy.
This indicates an IC to commensurate phase transition.
Such transitions are typically found to be first order because they involve the loss of a soft phase degree of freedom that exists in the IC state but is lost as the spin structure becomes commensurate and pinned to the lattice in the low-$T$ state. 
The transition also involves an abrupt drop in $\rho$ along the $b$-axis which is consistent with loss of scattering mechanisms associated with the phase degree of freedom in the IC state.
The transition is only visible as a weak inflexion in specific heat data \cite{Liang}, suggesting that the dominant modification at $T_2$ is a change in magnetic periodicity, which permits phase-locking the SDW to the lattice. 

For insights into the origin of incommensurability, we calculated the electronic band structure using the density functional in an all electron full potential linearized augmented planewave method with local orbitals. 
We used the WIEN2K \cite{wien2k} code with a generalized gradient approximation and spin orbit interactions. 
The corresponding Fermi surface is shown in Fig. 4 (a).
While the local chemistry of Fe is similar to that of other pnictides, the fermiology is distinct.
Consistent with the strip-like structure and the quasi-1D resistivity, the dominant features are sheets parallel to the $b$-axis.
As in the 122 pnictides, there are electron and hole tubes (Fig. 4(b)) near the corners of the Brillouin zone and $\Gamma$ point respectively.
The wave vector connecting the tubes, however, has no relation to magnetic ordering.
The dominant nesting instability appears to be associated with the planar Fermi sheets which are displaced from each other by distances close to $3{\bm b}^{\ast}/8$.
This indicates an itinerant description of magnetism and Fermi surface nesting may be relevant.
\begin{figure}[t!]
\includegraphics[width=155pt]{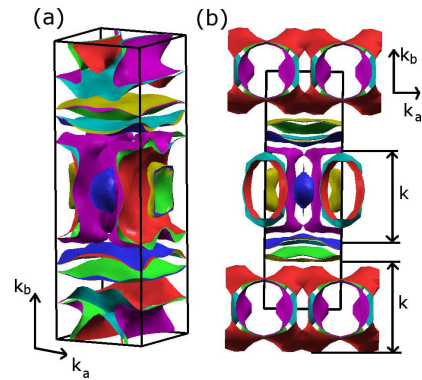}
\caption{(color online). (a) Fermi surface of CaFe$_4$As$_3$. Unlike in the 122 family, plane-like Fermi surfaces are dominant. (b) Viewing along $k_c$ reveals tube-like pockets as well as plane-like Fermi surfaces. The nesting vector $k=3/8$ is indicated.}
\end{figure}

We now turn to the detailed nature of the two magnetic phases at 50 K and 5 K using a single crystal oriented in ($hk0$) and ($0kl$) planes.
The observed intensities were compared to the intensities calculated for various irreducible representations (irreps) \cite{irreps} in Figs. 5.
The 50 K and 5 K spin structures corresponding to the best fits are shown in Figs. 1(c,d) and 1(e,f), respectively.
At 50 K, the data are consistent with magnetic moments ($\parallel b$) in irrep $\tau_4$.
Components within the $ac$ plane are allowed by symmetry but not supported by the data (0.04(13) $\mu_{\rm B}$).
The wave vector at this temperature is ($0,0.386(2),0$) and the spin structure may be described as a longitudinal SDW.
The best fit moment sizes are 2.1(1) $\mu_{\rm B}$ and 1.3(3) $\mu_{\rm B}$ per Fe$^{2+}$ and Fe$^{+}$ sites, respectively.

At 5 K, below the first order transition, ${\bm q}_{\rm m}$ is indistinguishable from ($0,3/8,0$), and the refinement indicates an $ac$ plane spin component described by irrep $\tau_1$ appears below $T_2$.
A full four-circle experiment to definitively determine the ordered spin structure awaits availability of larger single crystals.
The sequence of two phase transitions first to longitudinal IC order followed by the appearance of a transverse component with a different irrep also occurs in TbMnO$_3$ \cite{TbMnO3} and Ni$_3$V$_2$O$_8$ \cite{Ni3V2O8}, where it can be accounted for by a combination of competing exchange interactions and easy axis anisotropy as described by Ref. \cite{Nagamiya}. 
A longitudinal to transverse SDW transition is also often found in itinerant systems, Cr being a familiar example \cite{Cr}. 
As indicated by the enhanced low-$T$ intensity (Fig. 3(b)), the moment size increases below $T_2$ and the analysis yields 2.4(4) $\mu_{\rm B}$ per Fe$^{+}$ and 2.2(3) $\mu_{\rm B}$ per Fe$^{2+}$.
By far the largest increase occurs for the 5-fold coordinated Fe$^{+}$ site that links FeAs strips.
Overall, the ordered moment is substantially larger than 0.4-0.9 $\mu_{\rm B}$/Fe observed in the 1111 and 0.8-1.0 $\mu_{\rm B}$/Fe in the 122 systems \cite{rev}, suggesting stronger electron correlations. 
\begin{figure}[t!]
\includegraphics[width=240pt]{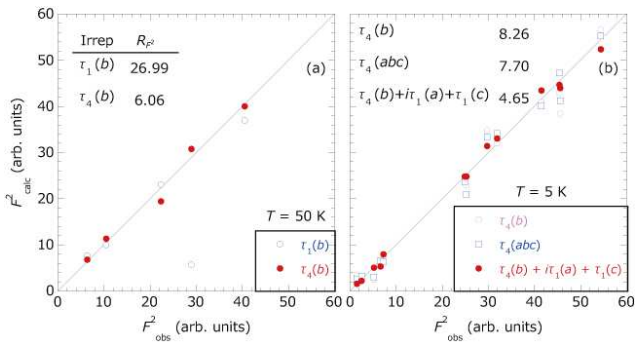}
\caption{(color online). Observed ($x$ axis) and calculated ($y$ axis) values for the squared magnetic structure factors within representative irreps at (a) $T=50$ K and (b) 5 K.}
\end{figure}

For a theory of magnetism in CaFe$_4$As$_3$, we performed first principles total energy calculations.
The antiferromagnetic (AF) state identified experimentally was found to have a lower energy than the ferromagnetic and paramagnetic states.
Analysis of the corresponding staggered magnetization reveals that while the fundamental periodicity prescribed by the supercell is 1/8, the dominant harmonic is 3/8 in good agreement with the experiments.
While the antiferromagnetism results in a calculated reduction of the density of states at the Fermi level by a factor of $\sim 2$, LDA predicts the low-$T$ magnetic state remains metallic as observed experimentally.

The structural similarity between the FeAs strips in CaFe$_4$As$_3$ and FeAs layers in the 1111, 122, 11 systems suggests analogies between their magnetic structures.
We shall describe these in the 2D centered square unit cell that holds two Fe atoms and has a lattice parameter $\sqrt{2}$ times the Fe-Fe spacing.
In an FeAs strip description of CaFe$_4$As$_3$, the $b$-axis corresponds to a lattice parameter in that cell.
The most common spin structure, AF2, has a wave vector (0.5,0.5) and corresponds to M-point Fermi surface nesting.
AF2 features nearest neighboring (nn) spins that are AF correlated in one direction and ferromagnetically correlated in the perpendicular direction.
It is called striped with stripes extending along nn bonds as favored by next nn exchange interactions.
AF2 is observed in the parent compounds of the 122 and 1111 materials and is associated with the spin fluctuations from which a spin resonance is formed in the superconducting state.
AF2 is to be distinguished from the conventional nn AF, which we denote AF1 with a wave vector (1,0) in the centered cell.
AF1 is also a striped structure but the stripes extend along next nn bond directions.
Favored by nn exchange interactions, AF1 is consistently observed in the parent compounds to cuprate superconductors but not so far in any compound related to iron pnictides.
Finally, AF3 has a wave vector (0.5,0) and is typically called double striped, stripes extending again along nn bonds in a structure that is favored by third nn exchange interactions.
AF3 and IC variants of it ($\delta$,0) with $\delta=0.3-$0.5 are observed in Fe$_{1+y}$Te$_{1-x}$Se$_x$ for small $x$ and finite $y$, respectively \cite{Bao}.
AF3 is analogous to the structure observed in CaFe$_4$As$_3$ where $\delta=3/8$ at low-$T$.
Indeed examining Fig. 1(d) one can appreciate the near double striped nature of the structure.
A similar periodicity is also reported for structurally related Fe$_{1.141}$Te \cite{Bao}.

It has been pointed out that thicker FeAs planes are associated with the AF3 structure while the AF2 structure occurs for thinner layers.
The layer thickness is 3.6 {\AA} in Fe$_{1.076}$Te \cite{Bao} with AF3 structure and 2.7-2.8 {\AA} in the 122 systems with the AF2 structure.
LDA calculations have suggested a critical layer thickness of 3.4 {\AA} for the transition from AF2 to AF3 related to the effect of changing bond angles on second and third nn exchange interactions \cite{height}.
In this context, it is interesting to note that the average layer thickness in CaFe$_4$As$_3$ ranges from 3.0-3.3 {\AA}; substantially exceeding that for the 122 and 1111 structures, though not quite reaching the inferred threshold for the AF3 structure.
To assess the relevance of competing exchange and local moment magnetism versus Fermi surface nesting and an itinerant description, doping studies should be useful as in metallic Cr \cite{Cr}.
A shift in the wave vector with carrier doping would indicate itinerant magnetism controlled by nesting. 

In summary we have examined FeAs magnetism in a unique strip geometry realized in CaFe$_4$As$_3$. 
A longitudinal SDW with IC modulation similar to observations in FeAs and Fe$_{1+y}$Te occurs below $T_{\rm N}=89.63(6)$ K.
A first order transition at $T_2=25.6(9)$ K yields a transverse component and locks the SDW to the lattice with a commensurate wave vector ($3{\bm b}^{\ast}/8$).
The calculated Fermi surface features sheets separated by a wave vector similar to that of the SDW, suggesting that Fermi surface nesting plays a role.
Alternatively the IC state could result from competing second and third nn exchange interactions in a localized spin picture.
Dramatic changes in $\rho$ through the lower-$T$ phase transition indicate proximity to $3d$ electron localization, a condition that appears favorable to high-$T_{\rm c}$ SC. 

Work at JHU is supported by the US DOE through DE-FG02-08ER46544.
Work at the NCNR is supported by the NSF under DMR-0454672 and the JSPS. 
Work at Rutgers Univ. and Rice Univ. is supported by DoD MURI Towards New \& Better High $T_{\rm c}$ Superconductors.

\end{document}